\documentclass[10pt]{iopart}%
\usepackage{times}
\usepackage{graphicx}
\newcommand{\eqref}[1]{(\ref{#1})}

\newcommand{\cat}[1]{\left|#1\right\rangle}
\newcommand{\roa}[1]{\left|#1\right\rangle\left\langle#1\right|}
\newcommand{\ro}[2]{\left|#1\right\rangle\left\langle#2\right|}
\newcommand{\aver}[1]{\left\langle #1\right\rangle}
\newcommand{\mess}[2]{\left\langle #1\left|#2\right|#1\right\rangle}
\newcommand{\brct}[2]{\left\langle#1\right|\left.#2\right\rangle}
\newcommand{\catt}[1]{\left|\left. #1 \right\rangle\right\rangle}
\newcommand{\catr}[1]{\left| #1 \right)}

\begin{document}

\title{Newly about reduction, non-locality and no-cloning}%

\author{Constantin V. Usenko}%

\begin{abstract}
	The problem of quantum state reduction in the process of measurement has attracted attention of almost everyone who created, developed or explained quantum physics to the students. Absence of a solution is the basis for the statement that the physical laws alone are not enough to describe the reduction of state of the measured particle.
	
	In the presented study, it is taken into account that the measurement of all properties of quantum particles can not be carried out in one physical process (the no-cloning theorem). For such a measurement several series of incompatible observations are needed, each series is to be carried out by means of a separate measuring instrument. Each event of measurement is a separate physical process; combination of different measurement events is performed by the Measurer - a subject capable of recognizing or not recognizing different events as similar depending on the purpose of the measurement, and capable as well of planning the measurements, of obtaining results and making conclusions.
	
	Several strategies for measuring the quantum state of photon polarization are considered; it is shown that modeling of a mixed state by a set of photon pure states is possible only if for this set in all the measurements the same measuring devices are used. It is also shown that the reduction of the quantum state is determined by the parameters of interaction of the measuring device with the measured particle. This fact removes the restrictions on the magnitudes of correlations of different observables, known as Bell's inequalities. 

\end{abstract}

\maketitle

\section{Introduction}

Reduction of state in the process of measuring the properties of quantum particles is the content of the last still unsolved fundamental problem of quantum mechanics. The problem is that, on the one hand, the state of the measured particle before the measurement is transformed during the measurement into one of several states determined by the properties of the measuring device. Quantum mechanics provides an algorithm for calculating the probability distribution of such a transformation. On the other hand, the result of each individual measurement event is a particular state, but it is not possible to specify exactly what the state is to be after the measurement. In the work \cite{epr} impossibility of accurate determination of the state in each individual measurement is characterized as incompleteness of quantum theory.

The state reduction is not specific for quantum particles only, a typical example of state reduction is found in the case of dice. The classical statistical physics explains the reduction by the fact that small differences in the initial conditions of a system comprising a measured object and a measuring device increase in the process of interaction. If the devices are very sensitive, the quantum properties of the system are substantial, so the problem is focused just around the reduction of quantum states.

The discussion of the reduction problem is present in almost all the courses of quantum mechanics, and practically all the quantum theory specialists are involved in it. Therefore, we can assume that explanation of the reduction goes beyond the physical picture of the Universe. One such possible explanation is presented in this paper. The author hopes for a fruitful discussion on the statement about absence of a physical resolution of the reduction problem, as well as on the involved notion about subjects of measurement and their purposeful activity.

The beginning of the history of the problem of reduction can be attributed to the Copenhagen interpretation of quantum mechanics laid down by Bohr and Heisenberg \cite{Heisenberg1927}. One of its consequences is the Heisenberg principle of uncertainty.

Almost complete description of the rules of reduction in physical processes is found in the monograph by John von Neumann \cite{vonneumann}. The physical formulation of the problem belongs to Einstein with Podolsky and Rosen \cite{epr} who have revealed contradictions between the assumption on locality and the possibility to assign the quantum particle exact values of all the observables. The Aspect's experiments  \cite{Aspect1,Aspect2}, along with the following experiments on checking the Bell's inequalities \cite{Bell}, have shown ineffectiveness of attempts to explain the reduction by the existence of hidden parameters of measured quantum particles.

The fundamental reason of absence of a purely physical explanation of the quantum particle state reduction phenomenon is the fact that the measurement for quantum particles is carried out not in one physical process, but in a set of different, though similar, physical processes. On the one hand, each measurement event is carried out in full accordance with the laws of physics which describe the interaction of the measured particle with the measuring device. On the other hand, acceptance of the set of events as similar and suitable for detection of statistical regularities of interaction of the measured particle with the measuring device makes need of participation of a Measurer --- a subject capable of planning and performing the required sequence of measurement events. The Measurer exists outside the physical system which comprises the measured particle and the measuring device.

The Measurer's task is to provide the required conditions: to prepare the desired state of the measured object; to choose the needed measuring devices; to register the obtained results. The presence of the stages of preparation and registration imposes restrictions on the time of interaction of the measured object with the measuring device.

The measurement consists of a series of measuring events; the series is characterized by the repetition rates of possible values, and those rates generate the estimate of the measured value probability  distribution. The accuracy of estimation depends on the number of repetitions, the standard error is inversely proportional to the root of the number of repetitions, so the required accuracy of measurement for a random quantity can be obtained only if the number of repetitions is not less than a certain value, called the measurement complexity \cite{DICE2012}.

The Measurer separates the measurement of a random quantity into a sequence or set of individual measurement events, adds to the measuring process comparison of the results of individual measurement events, and makes conclusions based on the analysis of the results. All those actions take place in accordance with the laws of physics, outside the measured particle and the measuring device, and should not affect the process of each individual measurement event.

The measurement of a quantum particle state further differs by the fact that there exist incompatible observables, complete information about the state of a particle can be obtained only as a result of measurement of several different observables (quantum tomography). For each observable a separate series of measurements is needed, therefore the complexity of measurement for a quantum particle is proportional to the number of independent incompatible observables; in the simplest case, a qubit, there are three such observables, for an N-level system the number of those is $N+1$, and for a harmonic oscillator or a free particle the required number of observables is infinite.

As a result, the task of planning a measurement for a quantum object is much more complicated compared to planning a classical measurement.

\section{Reduction of Polarized Light States}

The state reduction is here illustrated by polarized photons since for those the simplest --- two-dimensional --- state space is inherent, and the  devices for controlling polarization are well-known.

\subsection{Polarized Photons}

The pure polarization state is described by the complex polarization vector $\vec{e}_\mu$ with unit length, orthogonal to the direction of propagation. The set of polarization state vectors $\cat{\theta,\phi}$ is equivalent to a sphere (Poincare sphere) where each pair of diametrically opposite points corresponds to a pair of orthogonal basis vectors of the polarization state. For example, if a state represented by the vector $\cat{0} $ corresponds to the horizontal polarization, the state orthogonal to it, represented by the vector  $\cat{1} $, corresponds to the vertical photon polarization. The density matrix of an arbitrary pure polarization state $\roa{\theta,\phi}$ by means of the  Bloch vector with unit length $\vec{n}=\left\{\sin \theta\cos\phi,\sin \theta\sin\phi,\cos \theta\right\}$ can be represented by a linear combination of Pauli matrices
\begin{equation}\label{def_rho}
\hat{\rho}=\roa{\theta,\phi}=\frac{1}{2}\hat{I}+\frac{1}{2}\vec{n}\cdot\vec{\sigma}.
\end{equation}

The mixed polarization state is represented by a mix of a pair of orthogonal states $p_0\roa{0}+p_1  \roa{1} $ and is also described by density matrix \eqref{def_rho} in which the length of the Blokh vector is less than unity $\left|\vec{n}\right|= \left|p_0-p_1\right|$.

\begin{figure}[h]
\centering	\includegraphics[scale=1.0]{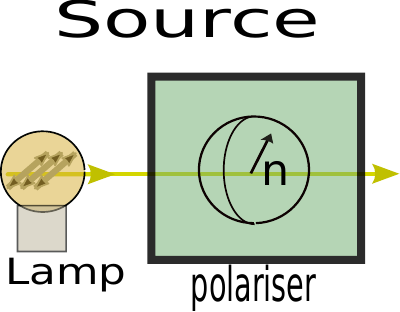}
	
	\caption{\label{fig_source}Source of polarized photons. 
		The lamp emits a photon with vertical polarization, the polarizer transforms polarization to the one characterized by the Bloch vector $\vec{n}$.}		
		
\end{figure}

Let us assume that we have a photon source that emits a photon with arbitrary given polarization. Let us also assume that the source emits photons  in pure states, since only at such an assumption the measurements the results of which have no statistical deviations can be carried out. Such a source can be a system shown in the picture \ref{fig_source}; this system consists of a source that emits only photons with vertical polarization, and a device that rotates polarization by a given angle $\theta $  and converts polarization to elliptic with a parameter $\phi$.

\begin{figure}[h]
\centering	\includegraphics[scale=1.0]{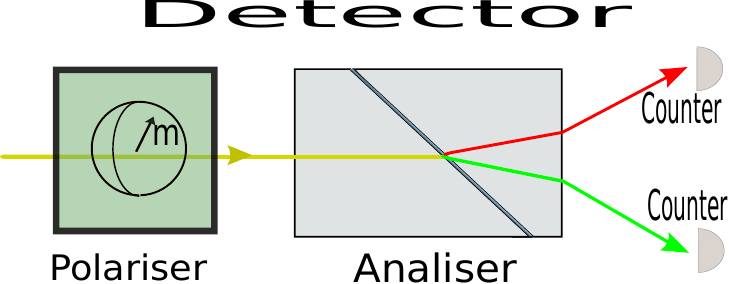}
	\caption{\label{fig_detector} Detector of polarized photons. 
		The polarizer transforms the photon polarization defined by the Bloch vector $\vec{m}$, to the vertical one. The analyzer splits the photon state vector into a superposition of vertically polarized and horizontally polarized states, the counter registers the photon.}		
		
\end{figure}

Measurement of the photon polarization is carried out by the detector shown in the figure \ref{fig_detector}. It consists of a polarizer that changes the orientation of the direction of polarization from one that is characterized by the Bloch vector $\vec{m}$ to vertical polarization (and any other polarization in same way); an analyzer that splits the beam into superposition of beams with vertical and horizontal polarizations; a pair of counters, separately for vertical and horizontal polarizations. 

The detector realizes the observable described by the matrix (operator)
\begin{equation}\label{def_obs}
\hat{M}=\vec{m}\cdot\vec{\sigma},\quad \vec{m}=\left\{\sin \theta_m\cos\phi_m,\sin \theta_m\sin\phi_m,\cos \theta_m\right\}
\end{equation}

The matrices of the observables characterized by different Bloch vectors do not commute, 
\begin{equation}
\left[M\left(\vec{m}\right)M\left(\vec{n}\right)\right]=iM\left(\vec{m}\times\vec{n}\right),
\end{equation}
corresponding observers can not be combined in one device. The existence of incompatible observables is the reason of impossibility of quantum state cloning, since for cloning the state is to be determined, and this leads to necessity of measurement of several, at least three incompatible observables. No-cloning property of quantum particles makes quantum measurements much more complex compared to classical ones.

The polarization state of the photon after the polarizer and before the analyzer is a superposition of states with vertical and horizontal polarizations

\begin{equation}\label{catPnm}
\cat{mes}=\sqrt{\frac{1+\vec{n}\cdot\vec{m}}{2}}\cat{1}+\sqrt{\frac{1-\vec{n}\cdot\vec{m}}{2}}\cat{0}.
\end{equation}

If the polarization state of the photon is pure, and the Bloch vector of the detector  polarization is the same as the Bloch vector of the photon polarization, the photon polarization state before the analyzer is the eigenstate 
 $\cat{1} $ of the analyzer matrix $
\hat{M}=\hat{\sigma}_3=\roa{1}-\roa{0} $, the measurement is a nondemolition one, and the statistical deviations of the repetitive measurements are absent.

Probability of registration with the upper or the lower in the figure \ref{fig_detector} counter for a photon prepared with polarization given by the Bloch vector $\vec{n}$ is
\begin{equation}\label{Pnm}
P_{v,h}\left(\vec{n},\vec{m}\right)=\frac{1}{2}\pm\frac{1}{2}\vec{n}\cdot\vec{m}.
\end{equation}

With this formula the possibilities of quantum theory as to prediction of the results of polarization state detection come to the end. The further adjustment of the measurement results requires additional notion about the phenomena occurring in each measurement event. 

\subsection{Photon-Detector Interaction}

Let's now take into account that the measuring device also changes its state in the process measurement, and these changes should take place in accordance with the laws of quantum mechanics. Since the device is a macroscopic system, its state is described by the density matrix $\rho_D$.

Before the measurement, the device and the photon are independent (and not correlated), so the common state of the 'photon+detector' system is a direct product of the density matrices $\rho_{in}=\rho_D\times \roa{mes}$. Interaction of the photon with the detector is described by the evolution matrix $\hat{U}\left(t\right)$ being the solution of the Schrodinger equation

\begin{equation}\label{Schreq}
i\frac{d}{dt}\hat{U}\left(t\right)=\left[\hat{H}\left(t\right)\hat{U}\left(t\right)\right].
\end{equation}

The Hamiltonian $\hat{H}\left(t\right)$ comprises the Hamiltonians of free detector $\hat{H}_D$ and the photon $\hat{H}_e$ and the interaction Hamiltonian

\begin{equation}\label{Hpd}
\hat{H}= \hat{H}_D+\hat{H}_e+\hat{H}_I,\quad \hat{H}_I=g\left(t\right)\hat{J}\times\left(\roa{1}-\roa{0}\right). 
\end{equation}

The measuring observable $\hat{J}$ does not commute with the Hamiltonian of the free detector, therefore in the 'detector+photon' system there are beats with alternation of states $\hat{\rho}_1\times\roa{1}$ ³ $\hat{\rho}_0\times\roa{0}$. The density matrix of the system is
\begin{equation}\label{rhoInt}
\begin{array}{l}
\hat{\rho}\left(t\right)=\hat{U}\left(t\right)\rho_{in}\hat{U}^+\left(t\right)=\\
p_1\left(t\right)\hat{\rho}_1\times\roa{1}+p_0\left(t\right)\hat{\rho}_0\times\roa{0}+\\\hat{\lambda}\left(t\right)\times\ro{0}{1}+\hat{\lambda}^+\left(t\right)\times\ro{1}{0}.
\end{array}
\end{equation}

Here $0\leq p_1\left(t\right)\leq 1$ and $p_0\left(t\right) =1-p_1\left(t\right)$ characterize the probability of the possible result of measurement, and $\hat{\lambda}\left(t\right) $  turns to zero only in the extreme cases, when the probabilities are equal to zero or one.

The Poincare recurrence theorem states that the density matrix will regularly approach these extremal cases. The art of designing detectors rests on making the time of transitions between extremal states much smaller than the time of being in vicinity of extreme states, with the last one proportional to probabilities \eqref{Pnm} of the respective results of measurement.

\subsection{Measurement Event Area}

The photon-detector interaction is always limited in time since it is preceded by preparation of the desired photon state, and registration of the measurement result must take place after that.

Each individual measurement event is restricted in time with the initial  $t_{i}$ and the end $t_{f}$ time instants. In the figure \ref{space_one} the space-time area for some measurement event $\mathcal{AREA}$ is represented. It is limited with the cone of the past $P_f$ with the apex at the end instant $t_{f}$ and the cone of the future $F_i$ with the apex in the initial instant $t_{i}$ and has a finite 4-D volume $\propto\left(t_{f}-t_{i}\right)^4 $.

\begin{figure}[h!]
\centering	\includegraphics[scale=0.25]{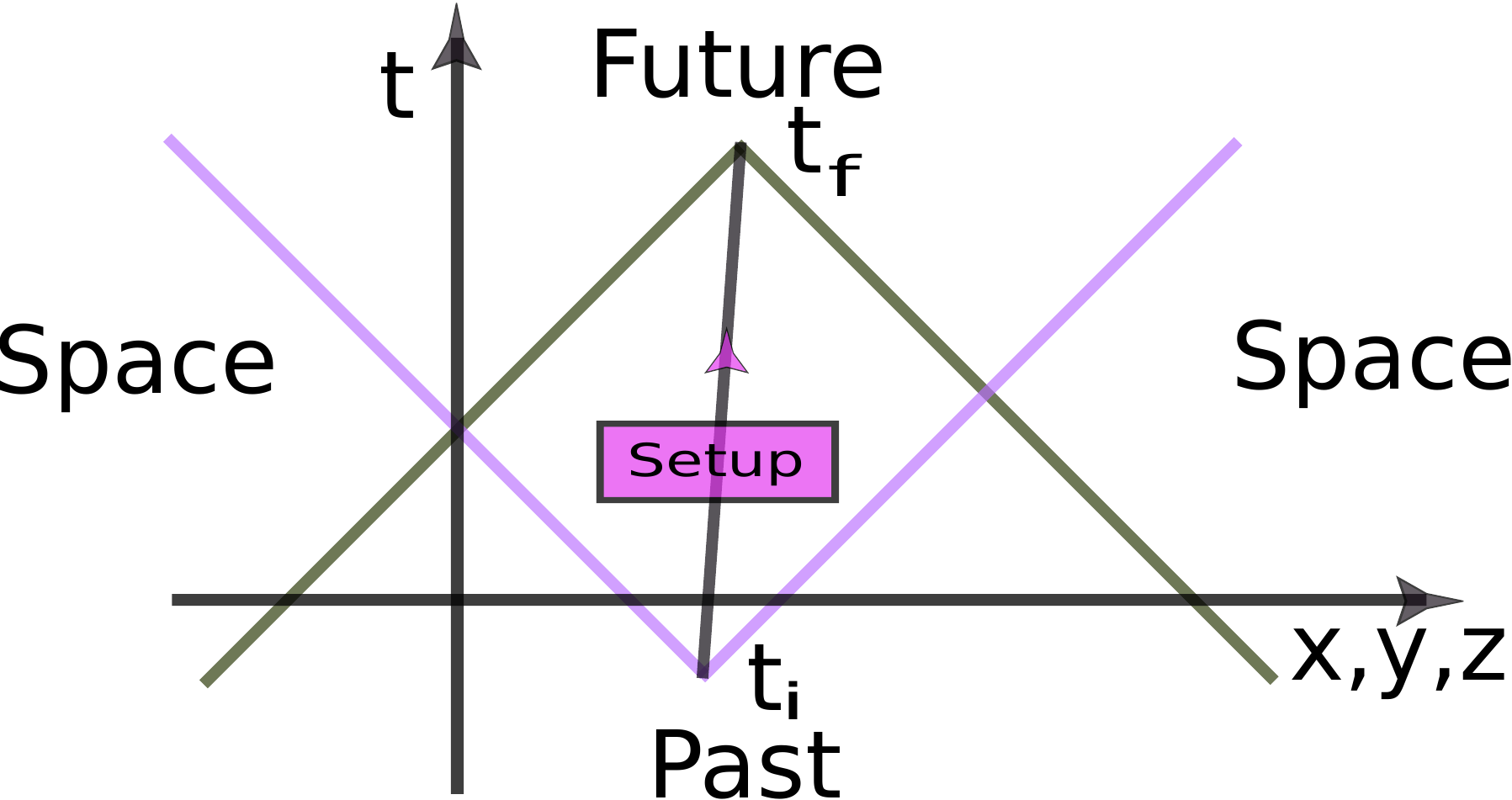}
	\caption{\label{space_one}Space-time area of measurement event.\\
	}\label{space_one}
\end{figure}

The detector in the process of measurement registers only the photons that fall on it after the initial instant $t_{i}$ and till the end instant $t_{f}$. All the registered photons are emitted in the space-time area bounded by the cone of the past $P_f$ with the apex in the end time instant $t_{f}$. The photons emitted in the cone $P_f$  outside the cone $F_i$, generate uncontrolled noise in the measured event.

All the events occurring outside the two cones form a spatially remote area of space-time $\mathcal{SPACE}$ that does not differ from infinitely remote. All the events taking place in the cone of the past $P_f$ with the apex in the point $t_{f}$ outside the cone of future $F_i$  with the apex in the po9int $t_{i}$ form the remote past $\mathcal{PAST}$ that does not differ from the infinitely remote past. All the events taking place in the cone of future  $F_i$  with the apex in the point $t_{i}$,  outside the cone of the past  $P_f$  with the apex in the point $t_{f}$, form the remote future  $\mathcal{FUTURE}$ that does not differ from infinitely remote future.

The fact that the measurement event is restricted in time and in space radically distinguishes the final state of the 'photon+detector' system  from the one predicted by the recurrence theorem. If the Poincare recurrence time is greater than the measurement time, no return occurs, the final state is determined by the parameters of the current state of the detector at the start time of the measurement and is virtually uncontrolled. A qualitatively built detector at the end of the measurement comes to one of the possible stable states $\hat{\rho}_1$ or $\hat{\rho}_0$. 

\subsection{Reduction}

The expression \eqref{Pnm} is valid only on average. If there are many photons, $N$, in the beam, part of those, in particular $N_1=\left(1+\vec{n}\cdot\vec{m}\right)/2$, is registered by the counter for vertical polarization, and $N_0=\left(1-\vec{n}\cdot\vec{m}\right)/2$ ---  with the counter for horizontal polarization. 

Quantum measurements correspond to a small number of photons, the most important is the case of single photons.

Since a photon is not divisible into parts between counters,  it is registered either by one or by another counter. In some cases the final polarization state of the photon is transformed into a state with vertical polarization, and in the other cases it is transformed into a state with horizontal polarization.

Before and after the end of the process of interaction of the detector with the photon, the intensity of interaction $g\left(t\right)$ in the expression \eqref{Hpd} turns to zero. Since the detector is a multiparticle complex system, the intrinsic parameters of the detector quickly vary in time, so the process of the detector-photon interaction depends not only on the duration of interaction, but also on the  detector state at the initial instant of interaction. Transformation of photon state into a state with the vector being the eigenvector of the measured observable is called the reduction of state. 


There exist two unitary matrices $U_v$, $U_h$, transforming the state \eqref{catPnm} into one or another reduced state:

\begin{equation}\label{Ureduct}
\begin{array}{lr}
U_v=\sqrt{\frac{1+\vec{n}\cdot\vec{m}}{2}}\left(\roa{1}+\roa{0}\right) &\\ +\sqrt{\frac{1-\vec{n}\cdot\vec{m}}{2}}\left(\ro{1}{0}-\ro{0}{1}\right),&\\
 & U_v\cat{mes}=\cat{1};\\
U_h=\sqrt{\frac{1-\vec{n}\cdot\vec{m}}{2}}\left(\roa{1}+\roa{0}\right) &\\ +\sqrt{\frac{1+\vec{n}\cdot\vec{m}}{2}}\left(-\ro{1}{0}+\ro{0}{1}\right),&\\
 & U_h\cat{mes}=\cat{0}.\\
\end{array}
\end{equation} 
For each measurement event the result of which is in transformation of the  photon polarization state into $\cat{1} $, as well as for each measurement event the result of which is in transformation of the photon polarization state into $\cat{0} $, the unitary matrix of photon polarization state transformation under effect of measuring device; transformation of the photon state takes place in accordance with the laws of quantum mechanics.  Only the expressions describing  the effect of the large system, the detector, on the small system, the photon, are different for different measurement events.

Thus, the reduction is caused by interaction of the photon with the detector; selection of one of two possible variants of the reduced state is determined by the hidden parameters of the detector. Repetitive return to initial state does not take place because of finite time of photon-detector interaction.

\section{Process of Measurement}

Measurement of photon polarization results in a random quantity with two possible values typically denoted as 0 and 1. The random quantity is characterized by probability distribution, and its determination requires a series of measurement events. The detector is designed so that the frequency distribution in a series generates an unbiased estimation of probability distribution. Since there are only two counters are in the detector \ref{fig_detector}, the frequency distribution, as well as the probability distribution, has only one independent value, for example, the frequency $\nu_1=N_1/\left(N_0+N_1\right) $, and respective probability is $p_1$. The second frequency $\nu_0=1-\nu_1$  completes the first frequency to one like the second probability  $p_0=1-p_1$ completes to one the first probability.

\subsection{Measurer}

The measurement differs from usual interaction between the measured particle and the measuring device by being carried out purposefully. The purpose of measurement may be to obtain information about the properties of a pure or mixed state of measured particle, or information on the state of the source of a particle, or to transfer information encoded by the properties of particle states from the source to the measuring device, or even to generate information, like in the methods of quantum key distribution.

The properties of the measurement events that make up the measurement are determined by a specific subject --- the Measurer who is not a part of the photon or the detector.

The Measurer, in accordance with the purpose of the measurement, organizes the measurement process by selecting the respective source and detector parameters for each measurement event. From the physical system the Measurer differs by capability to choose different scenarios in the same circumstances, depending on the purpose of the activity. In the model of polarized photons the Measurer can, at own discretion, choose for each measurement event the Bloch vector $\vec{n}$ of the initial state \eqref{def_rho} polarization of photon beam and the Bloch vector  $\vec{m}$, determining the observable \eqref{def_obs}. 

\subsection{Sequence of Measurement Events}

Organization of a set of measurement events can be serial, parallel or mixed. The serial organization excludes the possibility of interaction of photons belonging to different measurement events, though the measurement takes more time. The parallel organization can be more similar to a statistical ensemble, it requires efforts for spatial separation of photons.

In each measurement event the source $S$ emits a photon in a state specified by the Measurer; this state is described by a Bloch vector $\vec{n}_k $ specific for each measurement event. All the states of the sequence may be the same if the properties of one pure state are studied, or different, if this is prescribed by the purpose of the measurement. The result of operation of the source is described by a sequence of states
\begin{equation}\label{Sseria}
\mathcal{S} = \left\{\cat{\vec{n}_1},\ldots\cat{\vec{n}_k},\ldots\cat{\vec{n}_K}\right\}.
\end{equation}

In each measurement event the analyzer separates the states by the Bloch vector of the observable  $\vec{m}_k $ set by the Measurer for the event. The probability of registering polarized photons with the detector  $D_v$  are determined in each measurement event by the formula \eqref{Pnm} and form a sequence of predicted results of measurement events:
\begin{equation}\label{Pseria}
\mathcal{P} = \left\{\frac{1+ \vec{m}_1\cdot \vec{n}_1}{2},\ldots\frac{1+ \vec{m}_k\cdot \vec{n}_k}{2},\ldots\frac{1+ \vec{m}_K\cdot \vec{n}_K}{2}\right\}.
\end{equation}

Each measurement event results in registration of a photon with one of two counters,  $D_v$ or $D_h$. If we denote the fact of registering a photon by the counter  $D_v$ by the number 1, and with the counter  $D_h$ by the number -1, we get a numerical sequence of measurement results
\begin{equation}\label{Dseria}
\mathcal{D} = \left\{d_1,\ldots d_k,\ldots d_K\right\},\quad \forall d_k \in \left[-1,1\right].
\end{equation}

If the bit-tagging of the results is used, $D_v \mapsto 1,\ D_h\mapsto 0 $, we get a bit sequence
\begin{equation}\label{Bseria}
\mathcal{D} = \left\{b_1,\ldots b_k,\ldots b_K\right\},\quad \forall b_k \in \left[0,1\right].
\end{equation}

\subsection{Statistical Evaluation}

The scheduled by the Measurer sequence
$Q=\left\{\left[\vec{n}_1,\vec{m}_1\right],\ldots,\left[\vec{n}_k,\vec{m}_k\right],\ldots\right\} $  of pairs  $Q_k=\left[\vec{n}_k,\vec{m}_k\right] $ of the state polarization directions and detectors will be called the measurement strategy.

Let's now analyze the conditions under which the measurement strategy can be used for statistical estimation of the properties of the states emitted by the source.

\begin{enumerate}
	\item

	The simplest is the strategy in which all the prepared states are the same, $\forall k : \cat{\vec{n}_k}=\cat{\vec{n}} $; the same are as well the orientations of the detector, $\forall k : \cat{\vec{m}_k}=\cat{\vec{m}} $. We have a sequence of events with the same probabilities, and so we can calculate the frequency  $\nu_1=N_1/\left(N_0+N_1\right) $.

	Due to sameness of probabilities, the frequency gives an unbiased estimate of probability $p$, the variance $D=p\left(1-p\right)/{K}$ gives an estimate of the mean square error of estimated probability. Consequently, according to the results of the sequence of measurement events	
		we can estimate the probability p of the result and estimate the interval of possible deviations:	
	
	\begin{equation}\label{OneRes}
	p_d \approx \nu_d\pm \sqrt{\frac{\nu_d\left(1-\nu_d\right)}{K}}.
	\end{equation}
	\item 
	
	The strategy in which the prepared states are different and one observable is used, at first sight seems more complicated. To analyze the question on possibility of statistical evaluation let us consider the average value of the observable  \eqref{def_obs}.

	The result of measurement of one observable is $+1$, is the polarized photon is registered by the counter $D_v$, or $-1$, if the polarized photon is registered by the counter $D_h$. The set of results \eqref{Dseria} is the sequence of the values of the observable,		
	the probable value for a particular measurement event is equal to $\aver{\hat{M}}= \mess{\vec{n}}{\hat{M}}$,	
	and the mean value for all the measurements is determined by the formula
	\begin{equation}
	\aver{\hat{M}} = \frac{1}{K}\sum_{k=1}^{K}\mess{\vec{n}_k}{\hat{M}}.
	\end{equation}

Let us define the mean in sequence density matrix, and get the common expression for the mean value of the observable in the mixed state: 	
	\begin{equation}
	\rho=\frac{1}{K}\sum_{k=1}^{K}\roa{\vec{n}_k}.\quad \aver{\hat{M}} = \Tr{\rho\hat{M}}.
	\end{equation}
	
	This strategy makes it possible to simulate a mixed state with an arbitrary density matrix. Dispersion the density matrix component simulation does not exceed the value inverse to the number of members of the sequence. 	
		
	\item

	The following strategy is characterized by different observables and same state vectors. Let us suppose that for the measurement events with odd and even numbers the  observables  $\hat{M}_1=\roa{\vec{m}_1}-\roa{-\vec{m}_1}$ and $\hat{M}_2=\roa{\vec{m}_2}-\roa{-\vec{m}_2}$, respectively, are used. The odd and even members of sequence can be combined into two statistical subsequences with the same in each subsequence probabilities: 	
	\begin{equation}
\begin{array}{l}
	p_1^{\left\{1\right\}} = \frac{2}{K}\sum_{k=1}^{K/2}\mess{\vec{n}}{\roa{\vec{m}_1}}=\left|\brct{\vec{n}}{\vec{m}_1}\right|^2,\\
	p_1^{\left\{2\right\}}  = \frac{2}{K}\sum_{k=1}^{K/2}\mess{\vec{n}}{\roa{\vec{m}_2}}=\left|\brct{\vec{n}}{\vec{m}_2}\right|^2.
\end{array}	
\end{equation}

	Different statistical subsequences correspond to different observables. Each sequence apart from another one is characterized by an estimate of probability and its variance, as if two measurements of the simplest type have taken place separately. The course of measurement events resembles simultaneous execution of several programs by one processor, when at one instant of time only one program is executed, though does not come to the end naturally, but is forcibly terminated and replaced by another. Thus, this strategy carries out two independent measurements in quasi-parallel way. 	
	
	\item 
	
	Especially important is the strategy in which the measurement events differ both by the states and the observables.

	Let us consider as an example a specific chosen sequence of measurement events in which successively alternate  four states with Bloch vectors  $\vec{n}_1=0,0,1$, $\vec{n}_2=1,0,0$, $\vec{n}_3=0,0,1$, $\vec{n}_4=0,1,0$, and in turns two observables with the Bloch vectors $\vec{m}_1=0,0,1$, $\vec{m}_2=0,1,0$ are measured.

	The sequence of probabilities of the results is divided into four subsequences:	
	
	\begin{equation}
	\begin{array}{c}
	P_{1+4k}=\frac{1}{2}+\frac{1}{2}\vec{n}_1\cdot\vec{m}_1=1,\\ P_{2+4k}=\frac{1}{2}+\frac{1}{2}\vec{n}_2\cdot\vec{m}_2=\frac{1}{2},\\
	P_{3+4k}=\frac{1}{2}+\frac{1}{2}\vec{n}_3\cdot\vec{m}_1=\frac{1}{2},\\
	P_{4+4k}=\frac{1}{2}+\frac{1}{2}\vec{n}_4\cdot\vec{m}_2=1
	\end{array}.
	\end{equation}
	
	The average density matrix for the states registered with the detector of the first observable is defined as the weighted sum of density matrices of all the odd states	
	
	\begin{equation}
	\hat{\rho}^{\left\{1\right\}} =\frac{1}{2}\hat{1}+\frac{1}{2}\sigma_3=\roa{1}.
	\end{equation}
	It represents a pure state, since the odd states $\vec{n}_1=0,0,1$ and  $\vec{n}_3=0,0,1$ in this example are the same.	
	
	The average density matrix for the states registered with the detector of the second observable is defined as the weighted sum of density matrices of all the even states	
	\begin{equation}
\begin{array}{lr}
	\hat{\rho}^{\left\{2\right\}} =& \frac{1}{2}\left(\frac{1}{2}\hat{1}+\frac{1}{2}\sigma_1\right)+\frac{1}{2}\left(\frac{1}{2}\hat{1}+\frac{1}{2}\sigma_2\right)\\&=
	\left(\begin{array}{cc}
	\frac{1}{2}&\frac{1-i}{4}\\\frac{1+i}{4}&\frac{1}{2}
	\end{array}\right)
\end{array}
	\end{equation}
		It represents a mixture of states with weights $q_{1,2}=\left(1\pm 1/\sqrt{2}\right)/2 \approx [0.85 , 0.15]$.

	The dependence of the average density matrices on the choice of the conclusively necessarily denies the possibility of simulating the density matrix of an arbitrary state with a sequence of states, regardless of which observables are measured. 
 
 Density matrix can be simulated only by a sequence of measurement events where not only the states, but also the observables, are preselected.

\end{enumerate}

The main conclusion of this subsection is that the statistically significant sample is formed only by the measurement events in which the same observables are used. There are no reasons to assume that measurement events with different observables can be characterized by a common set of parameters, other than those knowingly laid down by the Measurer in the measurement strategy.

\section{Nonlocal Correlation}

The identity of the photons results in entanglement of states of an arbitrary pair of photons, therefore there is correlation between the results of measurements for the pair. The photons emitted by one source in different directions move at the light speed, and are thus separated with a space-like interval. Detectors of such photons do not carry out information exchange in the process of measurement. If the measurement results correlate, there is an erroneous impression that the information on state reduction is transferred from one detector to the other one in faster-than-light communication. Assumption about the existence of hidden parameters of the quantum state of entangled photons  \cite{epr} is aimed at restoring relativistic causality, so the results of Aspect's experiments \cite{Aspect1,Aspect2} on verification of the Bell inequalities have unexpectedly turned back the doubts on the  relativity theory. In this section we consider creation of information caused by the reduction of polarization states of an entangled photon pair.

\subsection{EPR-pair}

We consider a not polarized in average pair of photons (EPR-pair \cite{epr}) moving in opposite directions.

There is no polarization of the pair if the state of each of the photons is a superposition of states polarizations of which complement each other, and the polarization of the second photon complements the polarization of the first one.

Due to identity of the photons, the state vector of the pair is symmetric with respect to interchange of the photons. The set of such states forms a two-dimensional space. The basis of this space is formed by tow states each of which is a product of the states of spatial and polarization degrees of freedom:

\begin{equation}\label{degrees}
\begin{array}{rr}
\catt{S}=&
\frac{1}{\sqrt{2}}\left(\catr{\vec{p}}\times\catr{-\vec{p}}-\catr{-\vec{p}}\times\catr{\vec{p}}\right)\\
\otimes&
\frac{1}{\sqrt{2}}\left(\cat{\vec{n}}\times\cat{-\vec{n}}-\cat{-\vec{n}}\times\cat{\vec{n}}\right),
\\
\catt{T}=&
\frac{1}{\sqrt{2}}\left(\catr{\vec{p}}\times\catr{-\vec{p}}+\catr{-\vec{p}}\times\catr{\vec{p}}\right)\\
\otimes&
\frac{1}{\sqrt{2}}\left(\cat{\vec{n}}\times\cat{-\vec{n}}+\cat{-\vec{n}}\times\cat{\vec{n}}\right).
\end{array}
\end{equation}
Here ${\pm\vec{p}}$ stands for the parameters of the spatial state of photon pair, and $ \pm\vec{n} $ --- for those of the polarization state.

An arbitrary state of a not polarized in average photon pair is a superposition of the basic states:

\begin{equation}\label{param}
\catt{EPR}=\cos \theta \catt{S}+e^{i\phi}\sin \theta \catt{T}.
\end{equation}
The basic states \eqref{degrees} are characterized by independence of the results of measurement for polarization observables from the positions of the measuring devices, whereas the superposition states are characterized by correlation of the spatial and polarization degrees of freedom. This correlation can be programmed by the state Sender.

The specificity of a singlet state is that it looks the same in any basis:
\begin{equation}
\begin{array}{cl}
\cat{S}
=&\frac{1}{\sqrt{2}}\left(\cat{\vec{n}}\times\cat{-\vec{n}}-\cat{-\vec{n}}\times\cat{\vec{n}} \right) \\
=&
\frac{1}{\sqrt{2}}\left(\cat{\vec{m}}\times\cat{-\vec{m}}-\cat{-\vec{m}}\times\cat{\vec{m}} \right)|_{\forall \vec{m}}.
\end{array}
\end{equation}

The photons of the pair fall into detectors of polarization. The observables $\hat{M}_{l,r}=\vec{m}_{l,r}\cdot\vec{\sigma}$ for each detector are described by the Bloch vectors $\vec{m}_l$, $\vec{m}_r$. The mean values of both observables are zero, and their covariance is given by the expression:

\begin{equation}\label{EPR-cov}
D_{l,r}=\aver{\hat{M}_{l}\hat{M}_{r}}-\aver{\hat{M}_{l}}\aver{\hat{M}_{r}}=-\vec{m}_{l}\cdot\vec{m}_{r}.
\end{equation}

Measurement of polarization of one photon is accompanied by the pair state reduction. The reduced state of the second photon is pure. The respective Bloch vector $\vec{n}_2=-\vec{m}_1$ is opposite to the Bloch vector of the detector of the first photon.

\renewcommand{\labelenumi}{\theenumi)}

Thus, if the following conditions take place simultaneously:

\begin{enumerate}
	
	\item  polarization state of photon pair is a singlet one;

	\item  both polarization detectors have same Bloch vectors ($=\vec{m}_l=\pm\vec{m}_r$);

	\item  side effect or noise are absent;	
	
\end{enumerate}
there is to be observed 100-percent correlation or anticorrelation of the results of measurements of polarization.

If it is further assumed that the direction of motion of the photons does not change, the detectors of the polarization state of the entangled photon pair are to be separated by a space-like interval, and therefore can not interact with each other. However, you can imagine such a change in the direction of motion of the photons that they fall simultaneously to the pair of detectors located very closely. Due to absence of correlation of the spatial and polarization degrees of freedom, the results of measurement of polarization should not depend on the arrangement of the detectors.

In the other case, if the following conditions take place simultaneously:

\begin{enumerate}
	
	\item  polarization state of photon pair is a singlet one;

	\item  polarization detectors have orthogonal Bloch vectors ($\vec{m}_l\cdot\vec{m}_r=0$);	
	
	\item  side effect or noise are absent;
	
\end{enumerate}
correlation of the results of measurements of polarization is to be absent.

Since in singlet state the spatial and polarization degrees of freedom are independent, the listed predictions for polarization measurements  are independent from the arrangement of the detectors. It's enough that the detectors interact with the photons provided, if the carriers of the spatial part of the vectors are sufficiently located at the location of the detectors.  

\renewcommand{\labelenumi}{\theenumi.}

\subsection{Variants of organization of the  process of measurement}

Possibility of coordinated adjustment of detectors substantially depends on the purpose of measurement and the strategy of its achievement. 

It is rational to compare two different strategies:
\begin{enumerate}
	
	\item The first one is a research strategy with the purpose of verification the existence of a correlation between the results of measurements of polarization states of a singlet pair of photons. In this strategy the Measurer, before emission of photons, sends to the detectors  information on required (selected by the Measurer) values of the Bloch vector components, and after the interaction of photons with the detectors obtains the results of the current measurement event and can study the dependence of correlation on the angle between the  Bloch vectors of the detectors. The result of the measurement is the estimate of probability of deviation of correlation of the results from the expected 100-percent one.

	\item The second strategy aims at generating two copies of the secret key and is realized by two independent Measurers. Harmonization of the observables takes place at the stages of equipment preparation and adjustment. Deviations from coherence are compensated by the method of the key sifting   \cite{bennett96A}.	
	
\end{enumerate}

\subsubsection{Measurement of correlation of entangled pair}

\begin{figure}[!h]
\begin{center}
		\includegraphics[scale=.25]{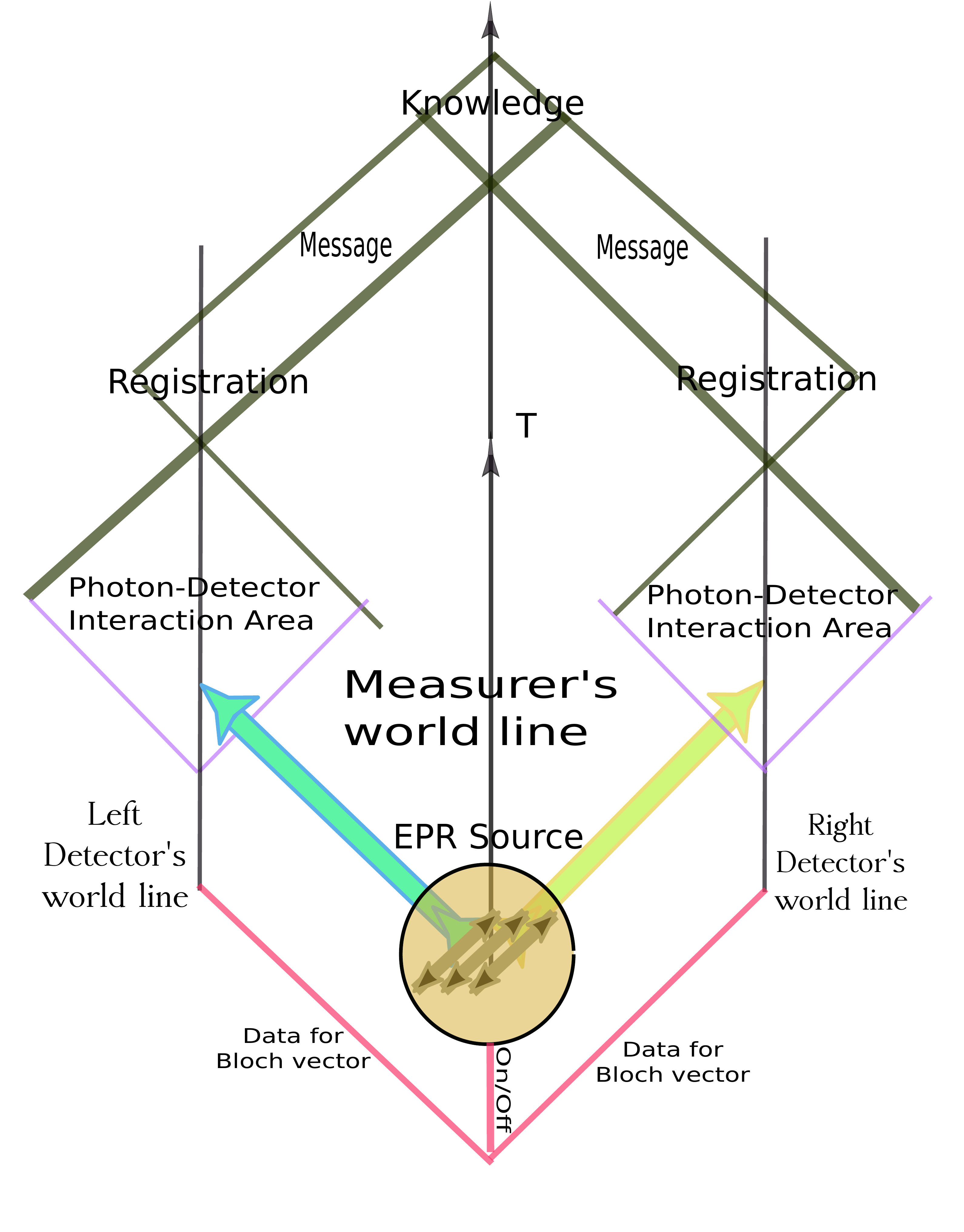}

	\caption{\label{fig_EPR} EPR-pair detection.	
		The Measurer determines the Bloch  vectors of the observables, as well as the instant of starting the measurement for the left and the right detectors, and activates the source of the entangled photon pair.		
		The EPR Source emits a pair of photons with singlet polarization state. The Detector Left and the Detector Right, separated by a space-like interval, register polarization of the photons. After the photon-detector interaction, the result is registered and a message with report about the result is sent. The messages from the right and left detectors are combined and information on correlation of the results is generated.}

\end{center}
\end{figure}

Measurements carried out to get knowledge about the correlation properties of an entangled photon pair require centralized coordination of processes that take place in the Source and in both Detectors. In the picture \ref{fig_EPR} the functional diagram of the the respective setup is shown.

The measurement event starts with the fact that the Measurer directs to the Detectors data determining the Bloch vectors  $\vec{m}_l$, $\vec{m}_r$ of the observables and indicating the instant of turning on the Detectors. After that, the Measurer switches on the Source of the entangled photon pairs. The photons fall onto the Detectors later compared to the data with determination of the Bloch vectors, so at the instant of arrival of the photons the Detectors are already set up and switched on. Further, the change of states of the Detectors takes place, this goes along with the reduction of polarization states of the photons of the pair. The detectors are switched off until the next measurement event, the detector states are registered, and a Report of the Results is sent to the Measurer. Having received both reports, the Measurer compares the results with the predictions of various hypotheses and evaluates the probability of the expected result.

Let us now suppose that the Source emits a pair of photons in a singlet state, and the Measurer selects the observables as complementary, with the opposite Bloch vectors   $\vec{m}_r=-\vec{m}_l=\vec{m}$. The basic hypothesis of the measurement is that the results of both Detectors should be the same, the alternative one assumes that those can be different with some probability $p_A$.

If in a series of measurement events with length  $ N$  the sub-series of outputs from the left
$\mathcal{D}_L=\left\{d_{1,L},\ldots, d_{N,L}\right\}$ 
and the right $\mathcal{D}_R=\left\{d_{1,R},\ldots, d_{N,R}\right\}$  detectors are same, $\left\{d_{1,R}=d_{1,L},\ldots, d_{N,R}=d_{N,L}\right\}$, we have for the alternative hypothesis upper probability limit the estimate  

\begin{equation}
p_A\leq P_A \left(N\right)= \frac{1}{N+1}.
\end{equation}

In particular, the first measurement event with the same results from both counters gives the upper limit of 1/2, and the psychological  threshold of 10\% is overcome only in 9 measurement events with the same results.

So, the measurement can confirm the 100-percent correlation of photon pair polarization with an error probability that does not exceed a given value $P$, if the length of the series of measurement events is not less than $1/P-1$.

The singlet states differ from the triplet ones by the fact that the 100-percent correlation of the results takes place for arbitrary orientation of the Bloch vector $\vec{m}$ of one of the observables, provided that the other observable is a complementary one, that is, it has orientation of the Bloch vector $-\vec{m}$. Verification of this statement requires realization of series of measurements with at least three complement pairs of detectors, and three more series are needed to confirm the absence of correlation if the Bloch vectors of detector pair are orthogonal. Thus, the required number of measurement events $N(P)$ confirming the correlation properties of the results of measurement of a singlet photon pair polarization is bounded below with the value

\begin{equation}
N(P) \geq N_s= \frac{6}{P}-1.
\end{equation}

The  measurement scheme presented here demonstrates that the absence of possibility of interaction of detectors does not lead to violation of causality, because, firstly, the possibility or impossibility of correlation is provided by the Measurer before the measurement, when the observables to be used in the detectors are determined; and secondly, correlation is determined not in each individual measurement event, but in the statistically significant series of measurement events.

Correlation between the observables of a singlet pair of polarized photons, on the one hand, completely corresponds to the expectations of quantum theory \cite{ AspectBell}, and on the other hand it does not deny the possible existence of hidden parameters \cite{epr}, with one substantial difference: the hidden parameters do not characterize the state of a photon pair as such, though characterize the interaction of photon pair with the measuring devices. Alternation of measuring devices is accompanied by alternation of the parameters, so correlation does not meet the restrictions known as Bell inequalities.

\subsubsection{Duplication of random sequence}

The practical application of correlation of polarization measurement results for an entangled photon pair in singlet state is based on the strategy of coordinating activities of two Measurers each of which operates one of two polarization state detectors. The results obtained by each Measurer form a random sequence. The purpose of coordination is to get the same sequences.

\begin{figure}[h]
\centering	\includegraphics[scale=.25]{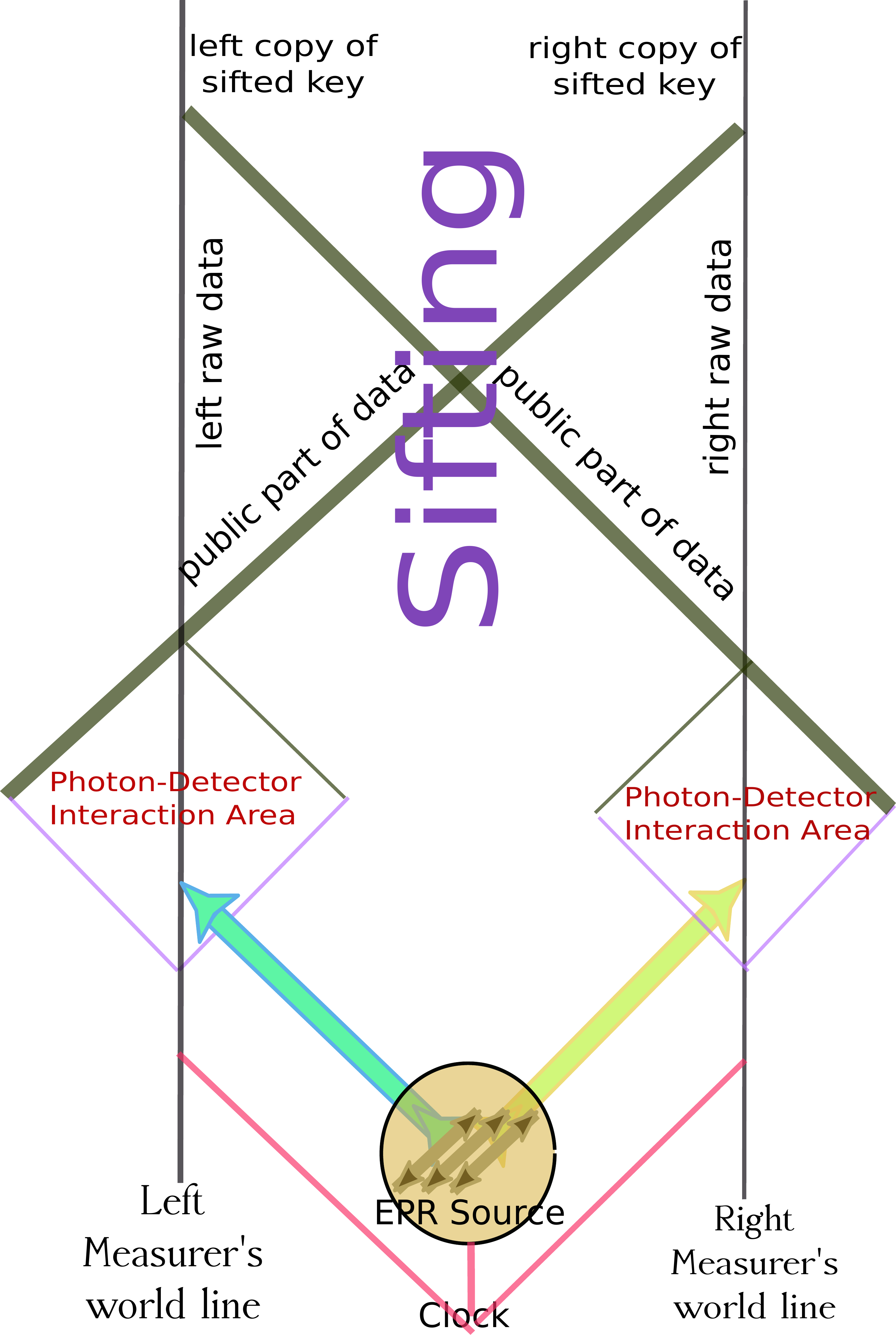}
	\caption{\label{fig_qkd} Quantum Key Distribution based on EPR-pair.
		The Bloch vectors of the observables are selected as antiparallel at designing the detectors. The instants of starting the measurement for the left and the right detectors and of turning on the source of entangled photon pair are determined by the clock combined with the source.		
			The EPR Source emits a pair of photons with singlet polarization state. The messages from the right and left detectors are sifted to remove errors, the same random sequences are left.}	
		
\end{figure}

Each Measurer in a routine measurement event randomly chooses one of two pre-coordinated variants of the Bloch vector directions $\vec{m}_k\leftarrow \vec{m}_{1,2}$.  By carrying out a measurement event, the Measurers compare the choices and get subsequences of the results  of measurements carried out by means of complementary detectors. Since the probability of differences in those sequences due to noise or purposeful third-party interference is not zero, a special sifting method is used \cite{bennett96A}.

Sifting makes it possible to delete from the sequences all the differences regardless of the origins of those, and makes of procedure of measuring polarization of an entangled photon pair in singlet state a practical protocol of quantum key distribution.

Sifting is needed due to the fact that theoretical computations for the probabilities of the results of measurements can only predict the possibility of generation of identical sequences, though do not guarantee the absence of effect of external factors that can cause uncontrolled change of the results.

 \section{Conclusion}
 
The Measurer as a subject setting up the conditions needed to obtain the measurement results not only combines different physical processes in one measurement, but also affects the properties of each individual measurement event. When preparing the measured object in the desired state, the Measurer bounds below the beginning of interaction of the measured object with the device. By registering the result of the measurement, the Measurer bounds above the time of the end of interaction of the measured object with the device. The consequence of these bounds is the possibility of designing measuring devices where there is no time to return to the initial state, as provided by the Poincare theorem. This fact provides the possibility of reducing the state of the measured object to one of the eigenstates of the measured observable.

 Measurement events are combined in a statistically meaningful sequence only if compatible measuring devices are used for those. Alternation of measuring devices in one measurement generates quasi-parallel measurements of several observables, with separate statistical subsequence for each observable.

 Correlation of the results of measurement of spatially separated observables of multiparticle states is a statistical effect that is observed only after statistical processing of the measurement results. The effect of unforeseen factors (noise, side interferences, etc.) can totally change the correlation, so practical application of correlation of the results of measurement is possible only after additional processing of the results (sifting for QKD protocols) aimed at removal of uncorrelated results.

Reduction of the measured state is the result of interaction with the measuring device. The direction of the reduction, that is, to which determined by the device state the measured object transits, is determined by the intrinsic parameters of the device just at the instant of measurement, and randomly varies from one measurement event to another. The intrinsic parameters vary with alternation of the devices, thus correlation of different observables is not restricted by the Bell inequalities, though it meets the requirements of quantum theory.

\end{document}